\documentclass[3p,times,procedia]{elsarticle}
\usepackage{nupha_ecrc}

\volume{00}

\firstpage{1}

\journalname{Nuclear Physics A}

\runauth{}

\jid{nupha}

\jnltitlelogo{Nuclear Physics A}

\usepackage{amssymb}

\usepackage{amsmath} 

\usepackage[figuresright]{rotating}

\usepackage{gensymb}  
\usepackage{floatrow} 

\newcommand{\AuAu}    {Au\,+\,Au collisions at 1.23\agev}   
\newcommand{\agev}    {\mbox{$A$~GeV}}               

\newcommand{\rb}[1]   {\mbox{\textrm{\scriptsize #1}}}

\newcommand{\pimin}   {\ensuremath{\pi^{-}}}

\newcommand{\rinv}    {\ensuremath{R_{\rb{inv}}}}

\newcommand{\qinv}    {\ensuremath{q_{\rb{inv}}}}
\newcommand{\qout}    {\ensuremath{q_{\rb{out}}}}
\newcommand{\qside}   {\ensuremath{q_{\rb{side}}}}
\newcommand{\qlong}   {\ensuremath{q_{\rb{long}}}}

\newcommand{\qoutsquare}   {\ensuremath{q^2_{\rb{out}}}}
\newcommand{\qsidesquare}   {\ensuremath{q^2_{\rb{side}}}}
\newcommand{\qlongsquare}   {\ensuremath{q^2_{\rb{long}}}}
\newcommand{\pt}             {\ensuremath{p_{\mathrm{t,}12}}}

\begin{document}

\begin{frontmatter}



\dochead{XXVIIIth International Conference on Ultrarelativistic Nucleus-Nucleus Collisions\\ (Quark Matter 2019)}

\title{Two-Pion Intensity Interferometry in Au+Au @ 1.23\agev}


\author{Robert Greifenhagen for the HADES collaboration}

\vspace{-20bp}

\address{Helmholtz-Zentrum Dresden-Rossendorf}

\begin{abstract}
High-statistics $\pimin\pimin$ HBT data for non-central \AuAu, measured with HADES at SIS18/GSI, are presented.
The three-dimensional emission source is studied in dependence on pair transverse momentum and centrality.
A tilt of the source relative to the beam axis is observed.
The spatial extension and the tilt magnitude of the source decrease with transverse momentum.
The spatial extension decreases and the tilt magnitude increases going from central to peripheral collisions.
The derived eccentricity perpendicular to the beam axis fits well to the initial nucleonic overlap region at high transverse momentum.
\vspace{0.2cm}
\end{abstract}


\end{frontmatter}

\pagestyle{plain}             

\section{Introduction}
\label{sec:introduction}

Two-particle intensity interferometry of hadrons is widely used to study the spatio-temporal size,
shape and evolution of their sources created in heavy-ion collisions or other reactions. 
At intermediate energies, various emission processes cause quite a number of origins of the measured particles. Therefore, the
intensity interferometry may provide additional information for the understanding of reaction mechanisms which finally
determine the particle emission sources.
Worth to mention is, that published HBT data in the $1A$ GeV energy regime are scarce, and that below $\sqrt{s_{NN}} \sim 10$ GeV only very
few analyses on azimuthally differential two-pion correlations have been performed.
The high statistics data set collected by HADES allows for the first time to perform a multi-differential femtoscopic pion analysis.
In this article we report azimuthally dependent two-pion intensity-interferometry results in non-central \AuAu ($\sqrt{s_{\mathrm{NN}}} = 2.41$ GeV) and derived geometrical quantities.
We shortly describe the HADES experiment, define the correlation function and introduce the used fit function. After this we present our observations and finally we summarize our results.

\section{The experiment}
\label{sec:experiment}
The data was taken with the High Acceptance Di-Electron Spectrometer (HADES) at the heavy-ion synchrotron SIS18 at GSI, Darmstadt. The setup of the HADES experiment is
described in detail in \cite{Agakishiev:2009am}. It is a fixed-target experiment with high angular acceptance ($18\degree-85\degree$),
operating at a high data rate. 
With a beam rate of (1.2-1.5)$\times 10^6$ ions/s, a trigger rate of up to 8 kHz and a target interaction probability of around 2\% within 557 hours
of beam taking more than two billion good events are collected for analysing in 45\% most central collisions.
Charged particles are identified measuring the curvature in a toroidal magnet field and relating it to the time-of-flight information.
The centrality determination of the events is based on the sum of measured charged particles compared to a Glauber Monte-Carlo (MC) calculation \cite{Miller:2007ri}.
Details can be found in \cite{Adamczewski-Musch:2017sdk}.
From this Glauber MC simulation we can furthermore deduce the initial nucleon eccentricities $\varepsilon_{\mathrm{initial}}$ for the chosen centrality windows.
The determination of the event plane angle $\phi_{\mathrm{EP}}$ is based on the detected charged projectile spectator fragments measured with a forward hodoscope. 

\section{The correlation function}
\label{sec:corr_function}

Experimentally the correlation is formed as a function of the momentum difference between the 
two particles of a given pair and quantified by taking the ratio of the yields of 'true' pairs 
($Y_\mathrm{true}$) and uncorrelated pairs ($Y_\mathrm{mix}$). $Y_\mathrm{true}$ is constructed  
from all particle pairs in the selected phase space interval from the same event.
$Y_\mathrm{mix}$ is generated by event mixing, where 
particle 1 and particle 2 are taken from different events.
Care was taken to mix particles from similar event classes in terms of multiplicity, vertex position and event plane angle.
The momentum difference is decomposed into three orthogonal components, \qout,~\qside~and \qlong, known as Bertsch-Pratt parametrization (suggested e.g. in \cite{Podgoretsky:1982xu}).
The particles forming a pair are boosted into the longitudinally comoving  
system, where the z-components of the momenta cancel each other, $p_{z_{1}}+p_{z_{2}}=0$. 
The experimental correlation function is given by  
\begin{equation}
C(\qout,\qside,\qlong) = {\cal N} \,\frac{Y_{\mathrm{true}}(\qout,\qside,\qlong)}{Y_{\mathrm{mix}}(\qout,\qside,\qlong)},  
\label{eq:def_exp_corr_fct_3dim}
\end{equation}
where $q_i=(p_\mathrm{1,\,i}-p_\mathrm{2,\,i})/2$ ($i$\,=\,'out',\,'side',\,'long') 
are the relative momentum 
components, and ${\cal N}$ is a normalization factor which is fixed by the requirement 
$C \rightarrow 1$ at large relative momenta, where the
correlation function is expected to flatten out at unity. The statistical errors of 
$C$ are dominated by those of the true yield, 
since the mixed yield is generated with much higher statistics.\\
The three-dimensional experimental correlation function is fitted with the function
\begin{equation}
C(\qout,\,\qside,\,\qlong) =  N \big[1-\lambda  + \lambda K_\mathrm{C}(\qinv,\rinv) C_\mathrm{qs}(\qout,\,\qside,\,\qlong)\big], 
\label{eq:pipi_fit_fct}
\end{equation} 
where
$C_\mathrm{qs} = 1+\exp{(-4 \sum_{i,\,j} q_i R_{ij} q_j)}$
represents the quantum-statistical part of the correlation function implementing a Gaussian source distribution, $\qinv= \lbrack \qoutsquare(1-\beta^2_{\mathrm{t}}) + \qsidesquare + \qlongsquare \rbrack^{1/2}$
is the invariant momentum difference with the transverse velocity $\beta_{\mathrm{t}} = 2k_{\mathrm{t}}/(E_1 + E_2)$ and \rinv~ the corresponding invariant HBT radius
from a corresponding 1-dimensional fit. 
The influence of the mutual Coulomb interaction in Eq.\,\eqref{eq:pipi_fit_fct} is separated from the Bose-Einstein part in the factor $K_\mathrm{C}$
by including in the fits the commonly used Coulomb correction
for extended sources by Sinyukov et al.\,\cite{Sinyukov:1998fc}.

\section{Results}
\label{sec:results}

\begin{figure}[ht]
\begin{center}
\includegraphics[width=1.0\linewidth]{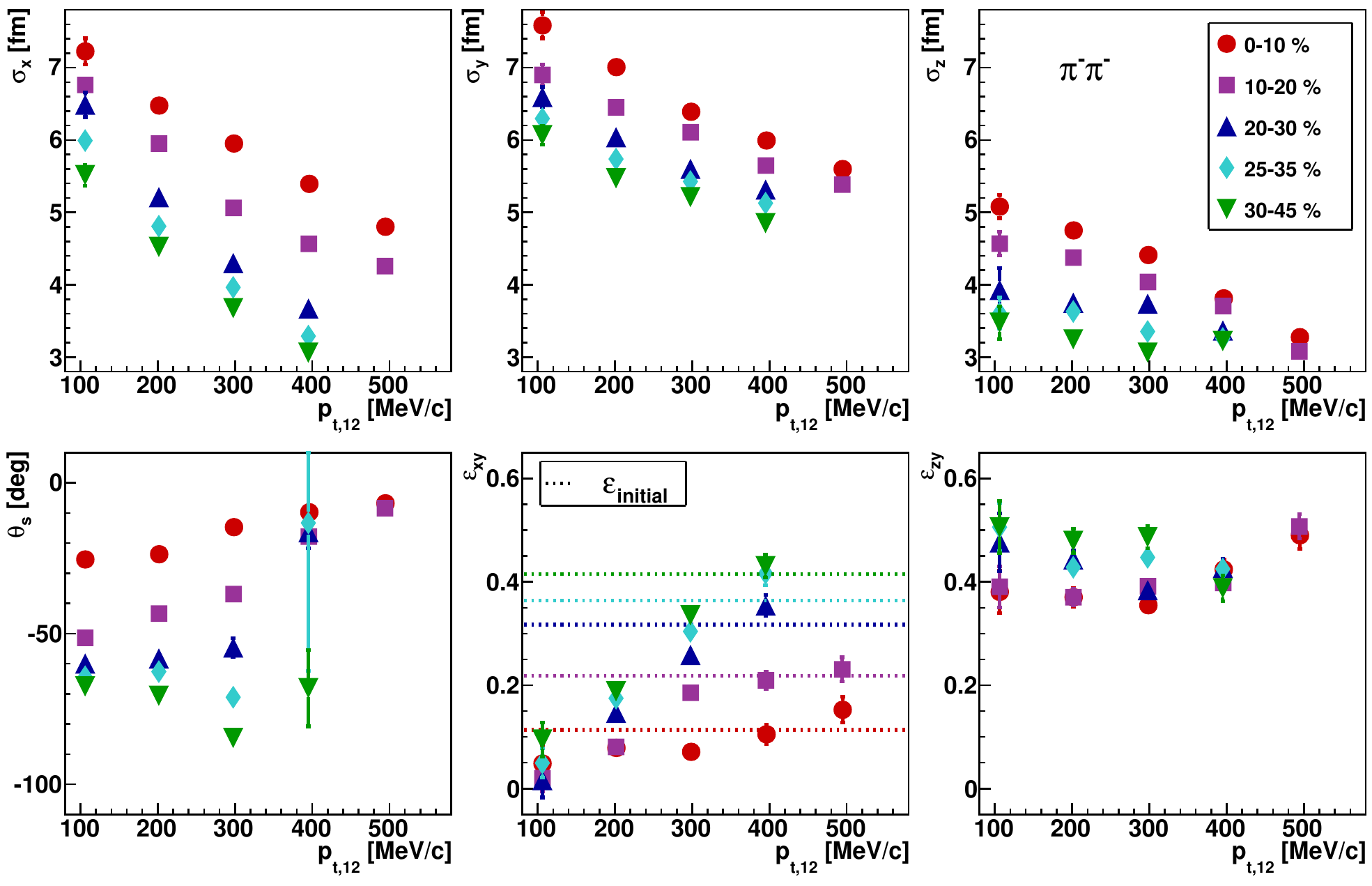} 
\caption{The spatial principal axes (upper panels), the tilt angle w.r.t. the beam axis in the 
reaction plane (Eq.\,(\ref{eq:tilt}), bottom left), the $xy$-eccentricity (bottom center)
and the $zy$-eccentricity (Eq.\,(\ref{eq:eps_xy}), bottom right) of the Gaussian emission ellipsoid
of $\pimin\pimin$ pairs as function of pair transverse momentum. Different symbols and colors belong to different classes of centrality
(red circles: 0-10\%, violet boxes: 10-20\%, blue up-pointing triangles: 20-30\%, cyan diamonds: 25-35\%, green down-pointing triangles: 30-45\%). 
Error bars include only statistical uncertainties. The dotted lines  
represent the initial nucleon eccentricities as derived from Glauber simulations.}
\label{fig:fig1}
\end{center}
\end{figure}

\begin{figure}[ht]
\floatbox[{\capbeside\thisfloatsetup{capbesideposition={right,center},capbesidewidth=3cm}}]{figure}[0.9\FBwidth]
{\caption{The $\pimin\pimin$ emission ellipsoid (final) $xy$-eccentricity (Eq.\,(\ref{eq:eps_xy})) as function 
of the initial nucleon eccentricity derived from Glauber simulations \cite{Miller:2007ri} 
for different pair transverse momentum classes. 
Error bars include only statistical uncertainties. The dashed line indicates 
$\epsilon_{\mathrm{final}} = \epsilon_{\mathrm{initial}}$.}
 \label{fig:fig2}}
{\includegraphics[width=1.0\linewidth]{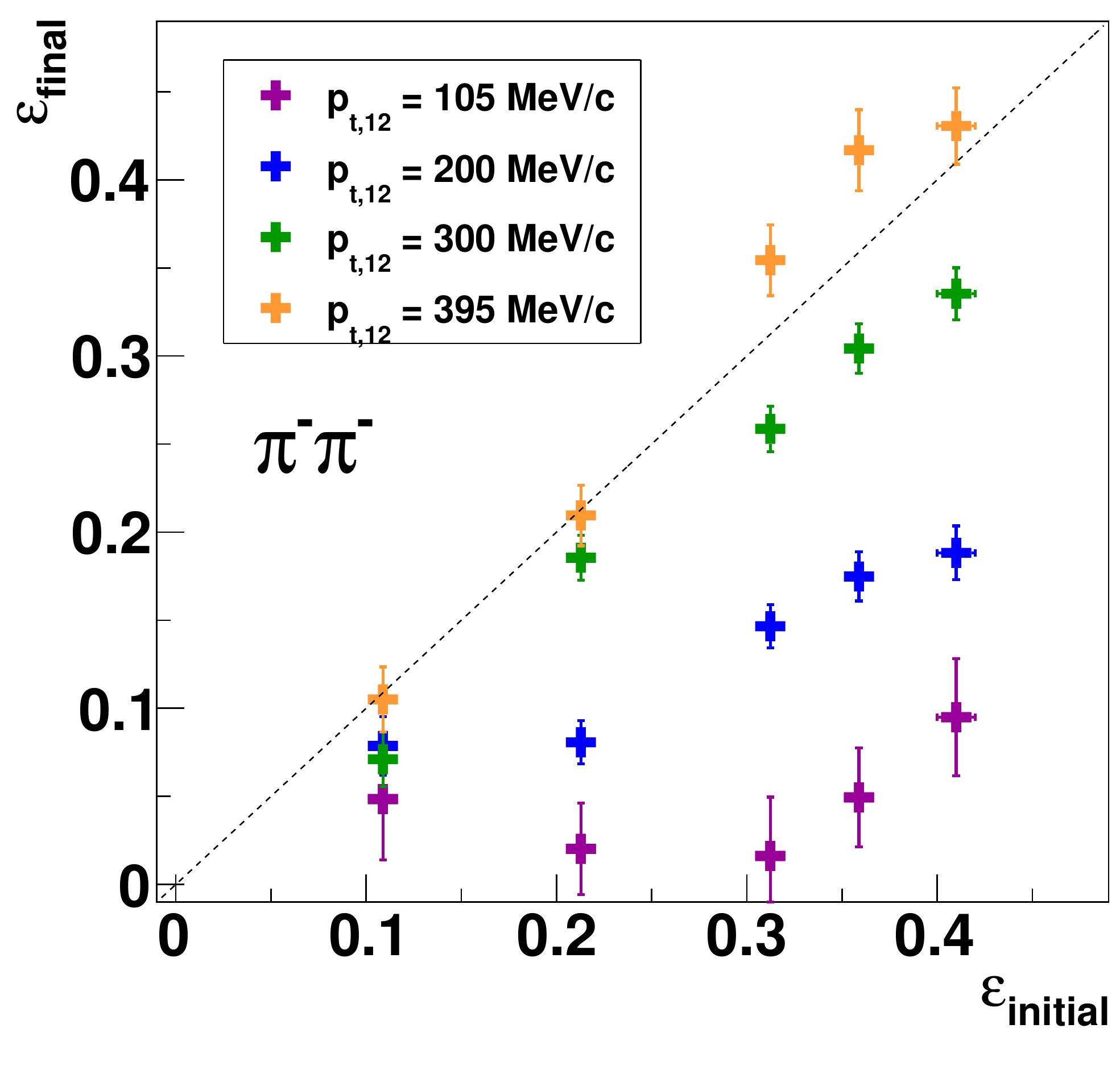}}
\end{figure}

The data set is divided into five classes of centrality (0-10\%, 10-20\%, 20-30\%, 25-35\%, 30-45\%) and intervals of pair transverse momentum
$\bold{p}_{\mathrm{t, 12}} = \bold{p}_{\mathrm{t, 1}} + \bold{p_{\mathrm{t, 2}}}$ with a width of
100 MeV/$c$ starting at 50 MeV/$c$, $k_{\mathrm{t}} = |\bold{p}_{\mathrm{t, 12}}|/2$.
Furthermore, the data set is divided into eight equal-sized intervals of azimuthal angle w.r.t. the event plane $\Phi = (\phi - \phi_{\mathrm{EP}})$.
The $\Phi$ dependence of the six HBT radius parameters extracted with Eq.\,\eqref{eq:pipi_fit_fct} can be found exemplary in \cite{Adamczewski-Musch:2019dlf}.
To account for the variations relative to the event plane the $\Phi$-dependence of the variables is fitted following \cite{Wiedemann:1997cr,Lisa:2000xj} with
\begin{align}
R^2_{\mathrm{out}} =\,&\,\frac{1}{2}(S_{11}+S_{22})+\frac{1}{2}(S_{22}-S_{11})\cos{(2\Phi)}+\beta^2_t S_{00}, 
~~R^2_{\mathrm{outlong}}=\,\,S_{13}\cos{(\Phi)}, \nonumber
\\
R^2_{\mathrm{side}}=\,&\,\frac{1}{2}(S_{11}+S_{22})+\frac{1}{2}(S_{22}-S_{11})\cos{(2\Phi)},
~~~~~~~~~~~~~~ \, R^2_{\mathrm{sidelong}}=\,\,-S_{13}\sin{(\Phi)}, \nonumber
\\
R^2_{\mathrm{long}}=\,&\,S_{33}+\beta^2_l S_{00},
~~~~~~~~~~~~~~~~~~~~~~~~~~~~~~~~~~~~~~~~~~~~~~~~~~~~~~~~~ \,R^2_{\mathrm{outside}}=\,\,\frac{1}{2}(S_{22}-S_{11})\sin{(2\Phi)},
\label{R2_azi} 
\end{align}
where the longitudinal pair velocity is $\beta_{\mathrm{l}} = 0$, and
$S_{\mu\nu}$ form a $4\times4$ matrix describing the 3+1 dimensional extension of the area of homogeneity in the $(xyz+t)$ system.
This set of fit functions accounts for the explicit $\Phi$ dependence arising from the
rotation of the osl-system relative to the RP-fixed $xyz$ coordinate system. The set of equations \eqref{R2_azi} fulfills several symmetry constraints of symmetric heavy-ion collisions \cite{Heinz:2002au}
and is performed simultaneously to all six variances of Eq.\,\eqref{eq:pipi_fit_fct} for each interval of $|\bold{p}_{\mathrm{t, 12}}|$ and centrality class separately.

The correction of both, the finite reaction-plane resolution and the finite azimuthal bin width, 
is performed by 
$R^{2,\mathrm{corr}}_{\mathrm{i,}n} = R^{2,\mathrm{meas}}_{\mathrm{i,}n} \, (n \, \Delta/2)/(F_n \sin(n \, \Delta/2))$, 
where $\Delta=\pi/4$ is the present bin width and the quantity 
$F_n$ represents the $n$-th event-plane resolution, determined by the sub-event method introduced in \cite{Ollitrault:1997di}.
$R^{2}_{\mathrm{i,}n}$ are the $n$-th coefficients of a Fourier expansion. One can assign the
required correction term by comparing the $S_{\mu\nu}$ elements in Eqs.\,\eqref{R2_azi} with the order of oscillation they appear with.
The used values of $F_1$ and $F_2$ can be found in \cite{Adamczewski-Musch:2019dlf}.
The spatial tilt angle in the $xz$-plane (modulo 90 degrees under exchange of $x$ and $z$) is given by
\begin{equation}
\theta_{\mathrm{s}}=\frac{1}{2}\tan^{-1}\left( \frac{2 S_{13}}{S_{33}-S_{11}}\right) .
\label{eq:tilt}
\end{equation}

Rotating $S_{\mu\nu}$ by the angle $-\theta_{\mathrm{s}}$, i.e. performing $G_y(\theta_s) S_{\mu\nu} G^{-1}_y(\theta_s)$ with the corresponding rotation matrix $G_y(\theta_s)$,
yields a diagonal tensor whose eigenvalues are the temporal and geometrical variances $\sigma^2_t, \sigma^2_x, \sigma^2_y, \sigma^2_z$.
The spatial variances are displayed in Fig.\,\ref{fig:fig1} (upper panels and lower right panel) as function of transverse momentum for \pimin\pimin.
The same figure shows the $k_{\mathrm{t}}$ dependence of $\theta_{\mathrm{s}}$ (lower left panel) and the $xy$($zy$)-eccentricity,
\begin{equation}
\varepsilon_{xy(zy)}=(\sigma_y^2-\sigma_{x(z)}^2) / (\sigma_y^2 + \sigma_{x(z)}^2),
\label{eq:eps_xy}
\end{equation}
in the lower middle (right) panel.
Due to the permutability of the $x$ and $z$ directions with corresponding shift in $\theta_\mathrm{s}$ by $\pm 90 \degree$, a variety of arrangements of the data points exists.
They were arranged such to obtain smooth trends for all observables in dependence of transverse momentum and centrality, smaller tilt angles for smaller impact parameters and vanishing
$\theta_\mathrm{s}$ at high values of \pt.

All semi-axes decrease with transverse momentum, similar to what was found for the azimuthally integrated results \cite{Adamczewski-Musch:2018owj}.
For all transverse momenta and all centralities, the 
deduced eccentricities represent an almond shape in the plane perpendicular to the beam 
direction.

In Fig.\,\ref{fig:fig2} the $xy$-eccentricity is related explicitly to the participant plane, 
as derived from Glauber simulations \cite{Miller:2007ri}.
For large transverse momenta the source eccentricity recovers the initial (nucleon) eccentricity, while at low values of \pt \
the shape becomes almost circular.

\section{Summary}
\label{sec:summary}
We presented high-statistics $\pimin\pimin$ HBT data for non-central \AuAu.
We studied the dependence on the pair transverse momentum and centrality.
We observe smooth trends of all derived geometrical quantities and a nice scaling of the $xy$-eccentricity at high \pt \ with the initial nucleonic participant distribution.





\bibliographystyle{elsarticle-num}
\bibliography{qm19_biblio}







\end{document}